\documentclass[conference]{IEEEtran}
\IEEEoverridecommandlockouts
\usepackage{cite}
\usepackage{amsmath,amssymb,amsfonts}
\usepackage{algorithmic}
\usepackage{graphicx}
\usepackage{textcomp}
\usepackage{booktabs}
\usepackage{xcolor}
\usepackage{hyperref}
\usepackage{multicol}
\usepackage{multirow}
\def\BibTeX{{\rm B\kern-.05em{\sc i\kern-.025em b}\kern-.08em
    T\kern-.1667em\lower.7ex\hbox{E}\kern-.125emX}}
\begin{document}

\title{MuCodec: Ultra Low-Bitrate Music Codec}

\author{

\IEEEauthorblockN{\scalebox{0.98}{\begin{tabular}{c}
        Yaoxun Xu$^{1,*}$\thanks{* Work performed during an internship at Tencent AI Lab.}, Hangting Chen$^{2,\dagger}$, Jianwei Yu$^{\dagger}$, Wei Tan$^{2}$, Rongzhi Gu$^{2}$, Shun Lei$^{1}$, Zhiwei Lin$^{1}$, Zhiyong Wu$^{1,3,\dagger}$\thanks{$\dagger$ Corresponding Author.}\thanks{This work has been submitted to the IEEE for possible publication. Copyright may be transferred without notice, after which this version may no longer be accessible.}
    \end{tabular}}}
\IEEEauthorblockA{$^1$ Shenzhen International Graduate School, Tsinghua University, Shenzhen, China\\
  $^2$ Tencent AI Lab 
  $^3$ The Chinese University of Hong Kong, Hong Kong SAR, China\\
xuyx22$@$mails.tsinghua.edu.cn erichtchen$@$tencent.com tomasyu@foxmail.com zywu$@$sz.tsinghua.edu.cn}
}

\maketitle

\begin{abstract}
Music codecs are a vital aspect of audio codec research, and ultra low-bitrate compression holds significant importance for music transmission and generation. Due to the complexity of music backgrounds and the richness of vocals, solely relying on modeling semantic or acoustic information cannot effectively reconstruct music with both vocals and backgrounds. To address this issue, we propose MuCodec, specifically targeting music compression and reconstruction tasks at ultra low bitrates. MuCodec employs MuEncoder to extract both acoustic and semantic features, discretizes them with RVQ, and obtains Mel-VAE features via flow-matching. The music is then reconstructed using a pre-trained MEL-VAE decoder and HiFi-GAN.  MuCodec can reconstruct high-fidelity music at ultra low (0.35kbps) or high bitrates (1.35kbps), achieving the best results to date in both subjective and objective metrics. Code and Demo: \href{https://xuyaoxun.github.io/MuCodec_demo/}{https://xuyaoxun.github.io/MuCodec\_demo/}.
\end{abstract}


\section{Introduction}
Music codecs\cite{codec1,codec2,codec3} are a crucial component in the field of audio codec\cite{audiocodec1,audiocodec2,audiocodec3} research. The significance of ultra low-bitrate compression lies in its potential applications, such as music transmission, where the bitrate of MP3\cite{casas2021mp3} is considerably high, and music generation\cite{copet2024musicgen,gao2024endtoendapproachchordconditionedsong,lei2024songcreatorlyricsbaseduniversalsong}, where short sequences are highly effective for language model construction. Furthermore, considering the diversity of background, sound events, and vocals in music, achieving high-fidelity reconstruction at ultra low bitrates would signify a substantial advancement in the field of universal audio generation.

Recent music compression techniques based on neural codecs\cite{nerualcodec1,nerualcodec2,nerualcodec3,nerualcodec4,nerualcodec5} attempt to compress music directly into discrete tokens. While discrete representations often yield higher compression densities, they inherently suffer from substantial information loss. To reconstruct a more accurate approximation of the original features from discrete tokens, a more robust representation and a stronger decoder are necessary. Common codecs like Encodec\cite{encodec} and Generative Adversarial Networks(GAN)-based methods\cite{gan1,gan2,gan3} exhibit limitations in achieving particularly low bitrates.

In recent years, some research and works have focused on using semantic modeling to represent musical characteristics and utilizing diffusion\cite{diffusion} for reconstruction, such as SemantiCodec\cite{liu2024semanticodec} and SEED-TTS\cite{anastassiou2024seed}. However, these models are not specifically designed for music-related tasks. Compared to speech tasks, music has a rich background, including instruments like piano and bass, and vocals that should be clearly discernible from the background music. Therefore, it is essential to consider both semantic and acoustic information; focusing solely on one aspect would compromise the overall perceptual quality of the reconstructed audio.

To address these challenges, we propose a flow-matching-based\cite{lipman2022flow} music codec MuCodec. MuCodec uses a specialized feature extractor, MuEncoder, based on the two key aspects of music: vocals and background. The MuEncoder features are then discretized using RVQ and employed as conditions for reconstructing Mel-VAE features via flow-matching. We reconstruct the Mel spectrogram by passing the Mel-VAE features through a pre-trained Mel-VAE decoder\cite{liu2024audioldm2}, and ultimately, the music is reconstructed using HiFi-GAN\cite{kong2020hifi}. Our contributions can be summarized as follows:
\begin{itemize}
    \item We propose MuCodec, which achieves the lowest bitrate to date while maintaining the highest-quality music reconstruction capabilities. 
    \item MuCodec employs MuEncoder as the feature extractor and Diffusion Transformer (DiT) \cite{dit} along with flow-matching-based method for fine-grained music modeling.
    \item Both subjective and objective experiments demonstrate that MuCodec achieves the best performance to date in music reconstruction tasks at both low and high bitrates.
\end{itemize}

\section{Method}
\begin{figure*}[htbp]
\includegraphics[width=2.0\columnwidth]{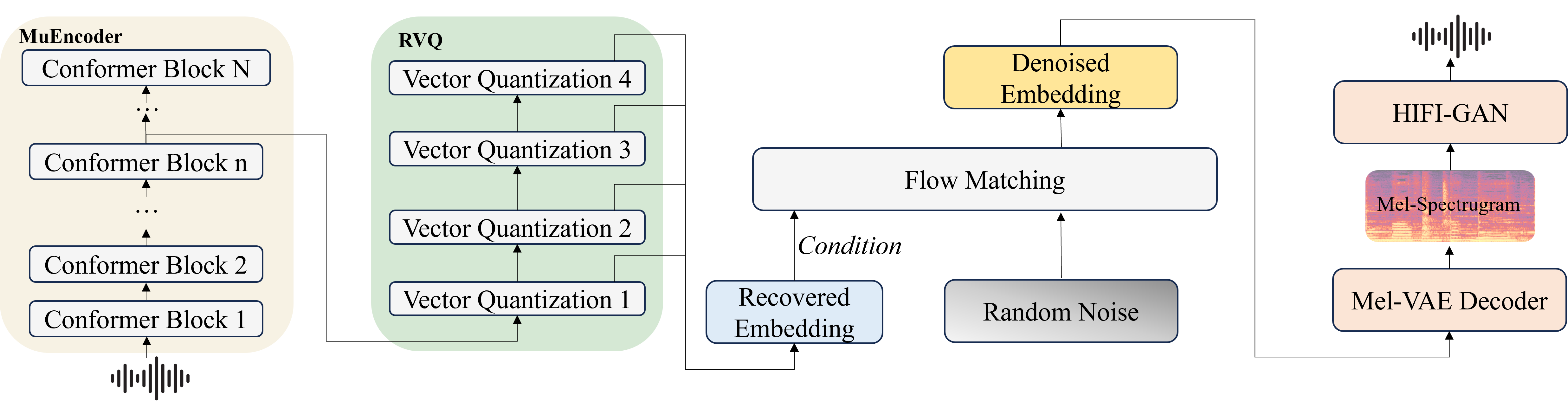}
\vspace{-4mm}
\caption{Framework of the proposed MuCodec.}
\vspace{-2mm}
\label{fig}
\end{figure*}

As illustrated in Fig. 1, MuCodec comprises MuEncoder, RVQ, a reconstruction model using flow-matching, Mel-VAE decoder, and HiFi-GAN. MuEncoder is a music extractor, primarily responsible for extracting both acoustic and semantic representations that better capture the characteristics of music. RVQ compresses the representations obtained from MuEncoder. The objective of flow-matching is to reconstruct low-bitrate discrete representations to obtain Mel-VAE features. Subsequently, the pretrained Mel-VAE decoder restores these features into a Mel spectrogram. Finally, the reconstructed music is obtained through a pretrained HiFi-GAN.

\subsection{MuEncoder}
Music reconstruction is more complex than speech or audio events, as it requires modeling both acoustic background and vocals. We design MuEncoder, composed of 13 stacked Conformer blocks, to extract acoustic and semantic features of background music and vocals.

To enable MuEncoder to extract both acoustic and semantic features, we implement a two-stage training process. In the first stage, we use the Mask Language Model constraint\cite{bestrq} to learn to predict masked regions based on unmasked speech signals, allowing MuEncoder to perceive contextual information and enhance representational capabilities. In the second stage, we introduce two constraints: reconstruction and lyrics recognition constraints. Reconstruction constraint aims to make extracted features closer to acoustic features, with two targets: restoring Mel spectrograms and predicting Constant-Q Transform (CQT)\cite{cqt} features. Lyrics recognition constraint ensures extracted features contain semantic information. These constraints enhance MuEncoder's feature extraction compatibility from both background music and vocal perspectives. 
\subsection{Residual Vector Quantization}
In MuCodec, we opt to use Residual Vector Quantization (RVQ) to discretize the MuEncoder features for its ability to compress representations through the residual process and provide more refined approximations using cascaded codebooks.

\subsection{Flow-Matching}
%For the reconstruction task, a direct and effective approach is to use Generative Adversarial Networks (GANs), which consist of a generator and a discriminator that progressively improve the generator's reconstruction quality through adversarial loss. However, GANs have limited modeling capabilities for ultra-low bitrate feature reconstruction, and it is challenging to solely use GANs to learn the process from discretized SSL features to audio. Therefore, 
MuCodec employs a flow-matching-based method for reconstruction, as it offers more stable training compared to GAN-based method and requires fewer training steps to achieve better results in ultra low-bitrate reconstruction task. Specifically, we use the discretized MuEncoder representations as a condition and perform finer-grained reconstruction through flow-matching with a Diffusion Transformer.

Instead of choosing the music or its Mel spectrogram as the flow-matching target due to their abundant and complex information, we predict the more manageable and information-rich Mel-VAE features for reconstruction. A pretrained Mel-VAE decoder serves as our Mel spectrogram generator, while a pre-trained HiFi-GAN functions as the music generator.
\begin{table*}[!h]
\label{tab:table1}
\centering
\caption{Objective experiment results on performance analysis. CodeBookSize is in the form of a x b, where a represents the number of codebooks and b represents the size of each codebook. SPK\_SIM refers to speaker similarity and the Visqol indicator is represented as x/y, with x and y denoting the Visqol scores for the left and right channels, respectively.}
\begin{tabular}{c|c|ccc|ccc}
\toprule[2pt]
                         &       Method           & CodeBookSize & Token Rate & kbps & VISQOL $\uparrow$ & SPK\_SIM $\uparrow$ & WER (\%) $\downarrow$   \\ \hline
    Origin music   & \textemdash & \textemdash & \textemdash & \textemdash & \textemdash & \textemdash & 10.92 \\ \hline
\multirow{5}{*}{\begin{tabular}[c]{@{}c@{}}Low-Bitrate\\ Scenario\end{tabular}} & DAC\cite{dac}+GAN              & 1 x 16384        & 25                                                             & 0.35 & 2.94/2.93 & 0.39                 & 131.80 \\
                         & MuEncoder+GAN        & 1 x 16384        & 25                                                             & 0.35 & 2.60/2.63 & 0.35                 & 97.86  \\
                         & MuEncoder+Diffusion  & 1 x 16384        & 25                                                             & 0.35 & 2.97/2.96 & 0.58                 & 89.31  \\
                         & SemantiCodec \cite{liu2024semanticodec}  & 1 x 32768            & 25                                                             & 0.375 & 1.92/1.92      & 0.52                 & 120.17 \\
                         & \textbf{MuEncoder+Flow-Matching (MuCodec)}       & 1 x 16384        & 25                                                             & 0.35 & \underline{3.09/3.08} & \underline{0.63}                 & \underline{68.37}  \\
                         & \textbf{MuCodec (200k)} & 1 x 16384        & 25                                                             & 0.35 & \textbf{3.19/3.20} &       \textbf{0.75 }              &  \textbf{40.81}    \\ \hline
\multirow{5}{*}{\begin{tabular}[c]{@{}c@{}}High-Bitrate\\ Scenario\end{tabular}}                 & DAC\cite{dac}+GAN              & 4 x 10000        & 25                                                             & 1.33 & 3.00/2.99 & 0.38                 & 137.51 \\
                         & MuEncoder+GAN        & 4 x 10000        & 25                                                             & 1.33 & 2.62/2.61 & 0.34                 & 62.59  \\
                         & MuEncoder+Diffusion  & 4 x 10000        & 25                                                             & 1.33 & 3.34/3.34 & 0.75                 & 43.36  \\
                         & SemantiCodec \cite{liu2024semanticodec}   & 1 x 16384            & 100                                                             & 1.40 & 1.96/1.96      & 0.68                 & 55.17  \\
                         & \textbf{MuEncoder+Flow-Matching (MuCodec) }       & 4 x 10000        & 25                                                             & 1.33 & \underline{3.30/3.30} & \underline{0.80}                 & \underline{34.19}  \\
                         & \textbf{MuCodec (200k)} & 4 x 10000        & 25                                                             & 1.33 & \textbf{3.43/3.42} &        \textbf{0.86}              &       \textbf{26.12}\\
\bottomrule[2pt]
\end{tabular}
\end{table*}

\subsection{Discussion}
\subsubsection{Disentangle}
In music reconstruction tasks, the two most important evaluation aspects are vocals and music background. To better verify the benefits of simultaneously focusing on these two features in music reconstruction tasks, we design comparative experiments to model these two aspects separately. Specifically, we choose pre-trained HuBERT\cite{hsu2021hubert} and MERT\cite{li2023mert} models to separately model vocals and music background. HuBERT typically contains richer semantic information, while MERT focuses more on acoustic features. 
\subsubsection{Scalability}
Although MuCodec is initially designed for music reconstruction tasks, it can also be easily applied to other types of audio without incorporating any additional training data, such as speech or acoustic events. MuCodec employs two constraints, one to enhance the background modeling of the audio itself and the other to strengthen the semantic modeling of vocals. As a result, MuCodec exhibits good performance in scenarios with pure vocals, pure background, or both vocals and background simultaneously. Our demo webpage exhibits the reconstruction results of different audio types and presents some other experimental outcomes.
\section{Experimental setup}
To train MuCodec, we utilize a large-scale internal music dataset of Chinese and English songs with a minimum 32kHz sampling rate. We segment the music into fixed 35.84-second lengths during training. For fairness, the test set comprises randomly selected 250 Chinese and 250 English song segments, each 20-30 seconds long with corresponding lyrics.

For the GAN-based method, we use a fully convolutional architecture following Descript Audio Codec(DAC)\cite{dac} encoder and decoder, changing the quantizer to RVQ. We match its model size to MuCodec. To further analyze, we also experiment with replacing its encoder directly with MuEncoder.

Considering GANs' weak reconstruction capabilities in low-bitrate scenarios, GAN-based method experiments are trained for 120k steps. In other cases, unless specifically stated, all test models train for 20k steps to demonstrate our approach's effectiveness within reasonable training time. 
Regarding SemantiCodec, we select two settings with bitrates similar to the high and low bitrates used in our experiments.

To better evaluate the performance of reconstructed music, we adopt both subjective and objective assessments. In subjective evaluations, we randomly select 5 Chinese and 5 English song clips as the test set and invite 10 professional participants to conduct a MUSHRA-inspired\cite{mushra} listening test. In objective evaluations, we choose two types of metrics corresponding to the two aspects of music: background and vocals. We use ViSQOL\cite{hines2015visqol} as an audio quality assessment metric.
Since the background can interfere with vocal evaluation, we separate the vocals and background of the generated music using a pre-trained separation model\cite{seperate}. We then calculate the similarity between the generated vocal part and the original vocal part with a pre-trained speaker similarity model\cite{speaker} and use Whisper-v2\cite{radford2023robust} to compute the Word Error Rate (WER) of the generated vocal part as the vocal clarity evaluation metric.

In the MuEncoder setting, we employ a 13-layer Conformer\cite{gulati2020conformer} model. We assign a weight of 1 to the music reconstruction loss and 0.2 to the lyrics recognition loss, which consists of both CTC Loss and RNN-T Loss.

In the RVQ setting, we design two configurations for high- and low-bitrate scenarios. For low-bitrate scenarios, we employ a single codebook with a size of 16,384 and a bitrate of 0.35kbps. Conversely, for the relatively high-bitrate scenarios, we use four codebooks, each with a size of 10,000, and achieve a bitrate of 1.33kbps. 
% we utilize a 32-dimensional codebook, set the quantizer dropout rate to 0, and allow a stale tolerance of 200.

In the flow-matching setting, we employ a 24-layer Transformer2d model\cite{transformer2d} for reconstruction, featuring an attention head dimension of 72, a norm epsilon of 1e-06, and 32 norm groups. We use ada norm single as the norm type and set the number of ada norm embeds to 1000. 
During the generation process, we utilize sampling via classifier-free guidance\cite{cfg}, specifically setting the guidance scale value to 1.5. 

During inference, we choose a denoising step size of 50 for flow-matching to balance reconstruction quality and computational efficiency. We use a pre-trained open-source Mel-VAE decoder and HiFi-GAN for both training and inference. Our experiments run on 8 40G-A100 GPUs with a batch size of 4.

\section{Result}
\subsection{Main Comparation}
% \usepackage{multirow}
% \usepackage{multirow}
% \usepackage{multirow}

In this experiment, we offer a thorough comparison of various prevalent reconstruction methods. MuCodec undergoes an in-depth comparative analysis from both objective and subjective assessments, with objective results in TABLE I.

First, it can be observed that DAC+GAN and MuEncoder+GAN method underperform in low-bitrate music reconstruction tasks, despite 120k training steps, which exceed other tasks.

Second, a difference between MuEncoder+Diffusion and MuCodec in low-bitrate music reconstruction tasks can be noticed. While MuEncoder+Diffusion outperforms GAN and MuEncoder+GAN, it falls short compared to MuCodec. This is because MuCodec employs the flow-matching method, which more directly and effectively models the noise-to-target distribution path compared to diffusion methods, achieving better results with fewer reconstruction steps.

Lastly, in the low-bitrate (0.35kbps) scenario, SemantiCodec's performance is subpar. Despite its state-of-the-art performance in acoustic event reconstruction, it lacks a dedicated design for music reconstruction tasks. Hence, its performance significantly decreases when handling more complex music reconstruction tasks compared to MuCodec. Furthermore, SemantiCodec only supports single-channel audio reconstruction at a 16k sampling rate, while MuCodec supports dual-channel audio at a 48k sampling rate, providing a greater advantage in music reconstruction.

At a higher bitrate (1.33kbps), MuCodec's performance continues to surpass other methods, showing the same trend as in the low-bitrate scenario. This demonstrates that MuCodec not only excels in low-bitrate scenarios but also delivers desirable results in high-bitrate music reconstruction tasks.

Moreover, we can observe from the table that when the training steps of MuCodec are increased to 200k, its performance improves further. However, a training step size of 20k already achieves a considerable level, highlighting MuCodec's robust compression and reconstruction capabilities.

\begin{figure}[htbp]
\centering
\vspace{-5mm}
\includegraphics[width=0.95\columnwidth]{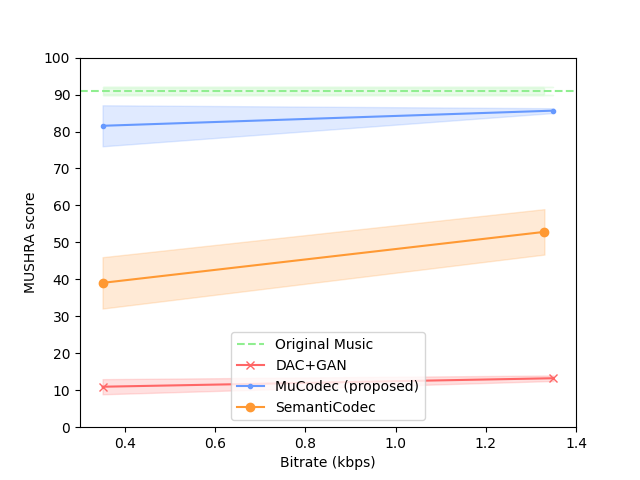}
\vspace{-3mm}
\caption{Listening test results on performance analysis.}
\vspace{-5mm}
\label{fig}
\end{figure}

Regarding the subjective results in Fig. 2, it is observed that the DAC+GAN method falls short in terms of audio quality at both low and high bitrates, indicating limited fine-grained modeling capability. In contrast, SemantiCodec shows a noticeable improvement over DAC+GAN method and performs better at high bitrates than low bitrates. However, despite its superior performance in acoustic event and speech reconstruction, the music reconstruction remains unsatisfactory, reflecting the challenges of music reconstruction tasks.

In comparison, our proposed MuCodec achieves excellent reconstruction results at both low and high bitrates, significantly outperforming SemantiCodec and DAC+GAN methods and closely resembling the original music. Moreover, the small difference between low and high bitrate MUSHRA scores suggests that MuCodec already attains a highly desirable reconstruction effect at extremely low bitrates. 

\subsection{Impact of Different MuEncoder Training Losses}

In this experiment, we evaluate the impact of MuEncoder on MuCodec under different loss conditions. Experiment \#1 uses only the Mask Language Model loss (MLM Loss). Experiment \#2 adds reconstruction loss (Recons Loss) to \#1, including Mel spectrogram and CQT feature reconstruction loss. Experiment \#3 incorporates lyrics recognition loss (ASR loss) based on \#2, with specific results shown in TABLE II.

\begin{table}[h]
\label{tab:table2}
\vspace{-4mm}
\caption{Experimental results on different MuEncoder training losses}
\centering
\begin{tabular}{l|c|ccc}
\toprule[2pt]
ID & MuEncoder Loss        & VISQOL $\uparrow$ & SPK\_SIM $\uparrow$ & WER (\%) $\downarrow$   \\ \hline
\#1 & MLM Loss & 2.70/2.71 & 0.587               & 84.41 \\
\#2 & \#1+Recons Loss  & 2.83/2.86 & 0.591               & 87.98 \\
\#3 & \#2+ASR Loss     & \textbf{3.09/3.08} & \textbf{0.631 }              & \textbf{68.37}\\
\bottomrule[2pt]
\end{tabular}
\end{table}

The results show that compared to \#1, ViSQOL and speaker similarity indicators improve in \#2 due to the additional reconstruction loss, while WER slightly decreases. This suggests that reconstruction loss enhances audio quality but has limited impact on vocal modeling. Comparing \#3 to \#2 reveals a significant WER reduction after adding recognition loss, benefiting vocal modeling and providing some support to ViSQOL and speaker similarity. This highlights that introducing reconstruction and recognition losses during training improves MuCodec's performance in music reconstruction tasks. 
\subsection{Influence of MuEncoder Layer Selection}

In this experiment, we evaluate MuCodec's performance under different MuEncoder layer conditions, specifically the 3rd, 7th, and 11th layers, with results in TABLE III.

\begin{table}[h]
\vspace{-3mm}
\caption{Experimental results on the MuEncoder layer selection}
\centering
\begin{tabular}{c|ccc}
\toprule[2pt]
MuEncoder Layer & VISQOL $\uparrow$ & SPK\_SIM $\uparrow$& WER (\%) $\downarrow$   \\ \hline
3           & \textbf{3.13/3.14} & \textbf{0.656}               & 76.65 \\
7           & 3.09/3.08 & 0.631               & 68.37 \\
11          & 2.92/2.92 & 0.618               & \textbf{63.46} \\
\bottomrule[2pt]
\end{tabular}
\end{table}

The results indicate that music reconstructed with lower MuEncoder layer features has better ViSQOL and speaker similarity indicators. As the number of MuEncoder layers increases, the reconstructed music quality decreases, while vocal clarity improves. This suggests that lower MuEncoder layers have stronger acoustic characteristics aiding in background music reconstruction, and higher layers contain more semantic features supporting vocal reconstruction. Therefore, in practice, the choice of MuEncoder layers needs to balance specific requirements, leading us to select the 7th layer as a balanced option in our experiments. 
\subsection{Validation Experiment for Disentangling Acoustic and Semantic Features}

In this experiment, we analyze the comparison between separate modeling and MuEncoder. We select the high-bitrate scenario in disentangle experiments to match MuCodec's high-bitrate case (1.33kbps). Separate HuBERT and MERT experiments use 4 codebooks, each with a size of 10,000, while joint modeling experiments with HuBERT and MERT allocate 2 codebooks for each model, each containing 10,000 elements. Specific results are detailed in TABLE IV.

\begin{table}[h]
\vspace{-3mm}
\caption{Experimental results on the feature extractor}
\centering
\scalebox{0.93}{
\begin{tabular}{c|c|ccc}

\toprule[2pt]
Feature Extractor         & Codebook & VISQOL $\uparrow$ & SPK\_SIM $\uparrow$& WER (\%) $\downarrow$   \\ \hline
HuBERT      & 4        & 2.54/2.54 & 0.312               & 80.78 \\
MERT        & 4        & 2.82/2.83 & 0.463               & 136.78  \\
HuBERT+MERT & 2+2      & \underline{3.12/3.14}  &   \underline{0.710}                  &  \underline{58.06}     \\
MuEncoder       & 4        & \textbf{3.30/3.30} & \textbf{0.804}               & \textbf{34.19} \\
\bottomrule[2pt]
\end{tabular}
}
\end{table}

It can be found that using HuBERT alone results in relatively low ViSQOL and speaker similarity, suggesting its inability to effectively model rich backgrounds. Using MERT alone improves audio quality and background but slightly decreases vocal clarity. Jointly modeling HuBERT and MERT features improves both background and vocal clarity without increasing bitrate. This suggests that jointly modeling vocals and background positively impacts overall music reconstruction but introduces additional computational complexity.

In contrast, using only MuEncoder yields better music reconstruction results than separate HuBERT+MERT and requires modeling only one type of feature. This makes it more suitable for modeling, prediction, and music generation tasks. 
% \subsection{Comparation on model size}
% \begin{table}[h]
% \caption{Table Type Styles}
% \centering
% \begin{tabular}{c|ccc}
% \toprule[2pt]
% DiT Params & VISQOL$\uparrow$ & SPK\_SIM$\uparrow$ & WER (\%)$\uparrow$  \\ \hline
% 266M       & 2.92/2.92 &    0.48                 & 80.70 \\
% 751M       & 3.09/3.08 & 0.63                & 68.37 \\
% \bottomrule[2pt]
% \end{tabular}
% \end{table}
% In this experiment, we compared the performance of MuCodec with different DiT parameters. The specific experimental results are shown in the table below. We adjusted the number of layers and attention heads in the Transformer2D, designing DiT models of different sizes.

% From the experimental results, it can be seen that as the DiT parameters increase, the performance of ViSQOL48K and SPKEAKER SIMILARITY indicators gradually improves, and the performance of the WER indicator also improves. This indicates that increasing DiT parameters during the training process helps to enhance MuCodec's performance in music reconstruction tasks.
\section{Conclusion}

To better address the challenge of ultra low-bitrate music reconstruction, we propose MuCodec, which achieves the lowest bitrate to date while maintaining excellent reconstruction music quality. MuCodec employs the MuEncoder feature extractor that considers both acoustic and semantic features of music, then the features are discretized using RVQ and finely reconstructed to Mel-VAE features via a flow-matching approach. The music is then reconstructed through a pretrained Mel-VAE decoder and HiFi-GAN. In both subjective and objective experiments, MuCodec significantly surpasses the current best results, realizing high-quality music reconstruction at an ultra low-bitrate scenario. 
\bibliographystyle{IEEEbib}
\bibliography{refs}

\end{document}